\documentclass{IEEEtran}
\usepackage{framed,multirow}
\usepackage{cite}
\usepackage{amsmath,amssymb,amsfonts}
\usepackage{algorithmic}
\usepackage{graphicx}
\usepackage{textcomp}
\usepackage{multirow}
\usepackage{comment}
\usepackage{tabu}
\usepackage{makecell}
\usepackage[colorlinks=true]{hyperref}

\def\citep{\cite}
\def\citet{\cite}
\newcommand{\bs}[1]{\boldsymbol{#1}}
\newcommand{\braces}[1]{\left\{{#1}\right\}}
\newcommand{\mR}{\mathbb{R}}

\begin{document}
\title{Knee Cartilage Defect Assessment by Graph Representation and Surface Convolution}
\author{
Zixu~Zhuang, 
Liping~Si, 
Sheng~Wang, 
Kai~Xuan, 
Xi~Ouyang, 
Yiqiang~Zhan,
Zhong~Xue, 
Lichi~Zhang, 
Dinggang~Shen,
Weiwu~Yao,
Qian~Wang

\thanks{Zixu Zhuang, Sheng Wang, and Xi Ouyang are with the School of Biomedical Engineering, Shanghai Jiao Tong University, Shanghai 200030, China. And they are also with the Shanghai United Imaging Intelligence Co., Ltd., Shanghai 200230, China.}
\thanks{Liping Si and Weiwu Yao are with the Department of Imaging, Tongren Hospital, Shanghai Jiao Tong University School of Medicine, Shanghai 200050, China.}
\thanks{Kai Xuan and Lichi Zhang are with the School of Biomedical Engineering, Shanghai Jiao Tong University, Shanghai 200030, China.}
\thanks{Yiqiang Zhan and Zhong Xue are with the Shanghai United Imaging Intelligence Co., Ltd., Shanghai 200230, China.}
\thanks{Dinggang Shen is with the School of Biomedical Engineering, ShanghaiTech University, Shanghai 201210, China. He is also with the Shanghai United Imaging Intelligence Co., Ltd., Shanghai 200230, China.}
\thanks{Qian Wang is with the School of Biomedical Engineering, ShanghaiTech University, Shanghai 201210, China (email: wangqian2@shanghaitech.edu.cn).}
}

\maketitle

\begin{abstract}
Knee osteoarthritis (OA) is the most common osteoarthritis and a leading cause of disability. Cartilage defects are regarded as major manifestations of knee OA, which are visible by magnetic resonance imaging (MRI). Thus early detection and assessment for knee cartilage defects are important for protecting patients from knee OA. In this way, many attempts have been made on knee cartilage defect assessment by applying convolutional neural networks (CNNs) to knee MRI. However, the physiologic characteristics of the cartilage may hinder such efforts: the cartilage is a thin curved layer, implying that only a small portion of voxels in knee MRI can contribute to the cartilage defect assessment; heterogeneous scanning protocols further challenge the feasibility of the CNNs in clinical practice; the CNN-based knee cartilage evaluation results lack interpretability. To address these challenges, we model the cartilages structure and appearance from knee MRI into a graph representation, which is capable of handling highly diverse clinical data. Then, guided by the cartilage graph representation, we design a non-Euclidean deep learning network with the self-attention mechanism, to extract cartilage features in the local and global, and to derive the final assessment with a visualized result. Our comprehensive experiments show that the proposed method yields superior performance in knee cartilage defect assessment, plus its convenient 3D visualization for interpretability.
\end{abstract}


\section{Introduction}
\IEEEPARstart{K}{nee} osteoarthritis (OA) is the most common osteoarthritis. More than 240 million people are suffering from knee OA in the world~\cite{arden2021non}. 
If not treated in time, the incidence and severity of OA will increase with aging and eventually lead to disability at the end stage~\cite{vos2016global, si2020knee}.
So far, joint arthroplasty has been the only effective treatment to cure knee OA at the end stage, although prostheses can only provide a restricted exercise ability and its lifespan is limited~\cite{glyn2015osteoarthritis}. 
Many researchers thus focus on the diagnosis and prevention in the early stage of knee OA~\cite{scanzello2012role,malfait2013towards}. 

\textbf{\begin{figure}[t]
\centering
\includegraphics[width=\columnwidth]{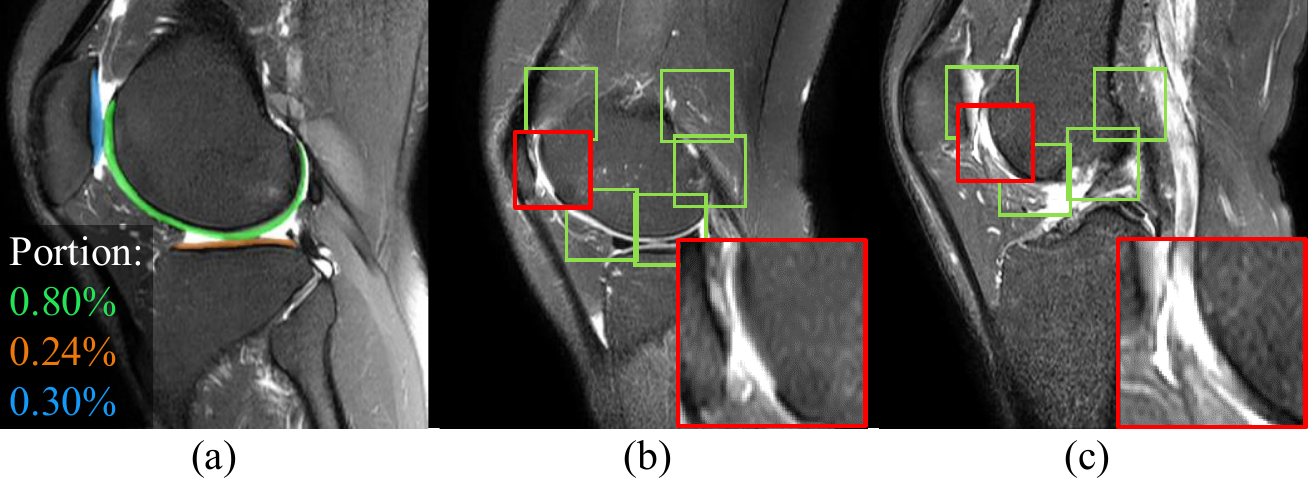}
\caption{
There are many challenges for deep networks to assess knee cartilage defects: (a) \textbf{Information Dilution}: the cartilages (labeled in colors) occupy only a small portion of voxels, compared to the large background, in knee MRIs; (b-c) \textbf{Structure Disidentification}: along with the curved cartilage, the local appearance of individual patches can be highly confusing to determine whether the patches contain defects. For example, (b) and (c) both show the discontinuity of femoral cartilages. However, (b) is the femoral cartilage with a serious defect, and (c) is normal at the end zone of the femoral cartilage.
}
\label{fig_intro}
\end{figure}}

Although the cause and pathological changes of knee OA are comprehensive, knee cartilage defects and changes in the adjacent bones are regarded as typical manifestations of knee OA~\cite{arden2006osteoarthritis}.
In knee OA diagnosis, magnetic resonance imaging (MRI), as a non-invasive and radiation-free examination, shows superiority for comprehensive knee pathology assessment, because of its better contrast for soft tissues (e.g., cartilages and edema areas) compared to X-ray and computed tomography~\cite{glyn2015osteoarthritis, blumenkrantz2007quantitative}.
However, the cartilage defect assessment results might vary among radiologists. 
To eliminate this variation and quantify the severity of knee cartilage defects, several grading systems are adopted, e.g., MRI Osteoarthritis Knee Score (OKS)~\cite{hunter2011evolution} and Whole-Organ MRI Score (WORMS)~\cite{alizai2014cartilage}. 
These grading systems also allow computer-aided diagnosis~(CAD) to be introduced into cartilage defect assessment, providing accurate and detailed results and improving the efficiency of diagnosis.

There are multiple CNN-based types of research focusing on the automatic assessment of cartilage defects, which are especially inspired by the recent leap of deep learning and may potentially contribute to the intelligent diagnosis of knee diseases~\cite{liu2018deep, pedoia20193d, bien2018deeplearningassisted, guida2021knee, liu2019fully, huo2020self, merkely2021improved}.  Typically these attempts can be categorized into the following three ways.
\begin{itemize}
    \item Patch-based methods: 
    Liu et al.~\cite{liu2018deep} cropped patches along with the femoral cartilage. Then they used the encoder in the U-Net architecture to help classify the cartilage defects in the patches.
    \item Slice-based methods: 
    Huo et al.~\cite{huo2020self} provided a self-ensembling framework with dual consistency loss, such that the trained CNNs can better concentrate their attention on OA-affected cartilage regions in the MRI slice. 
    \item Subject-based methods:
    Guida et al.~\cite{guida2021knee} used 3D CNNs to deal with the whole knee MRI. They claimed that 3D CNNs can extract the 3D features and thus perform better than the slice-based methods.
\end{itemize}

\begin{figure*}[!htbp]
\centering
\includegraphics[width=\textwidth]{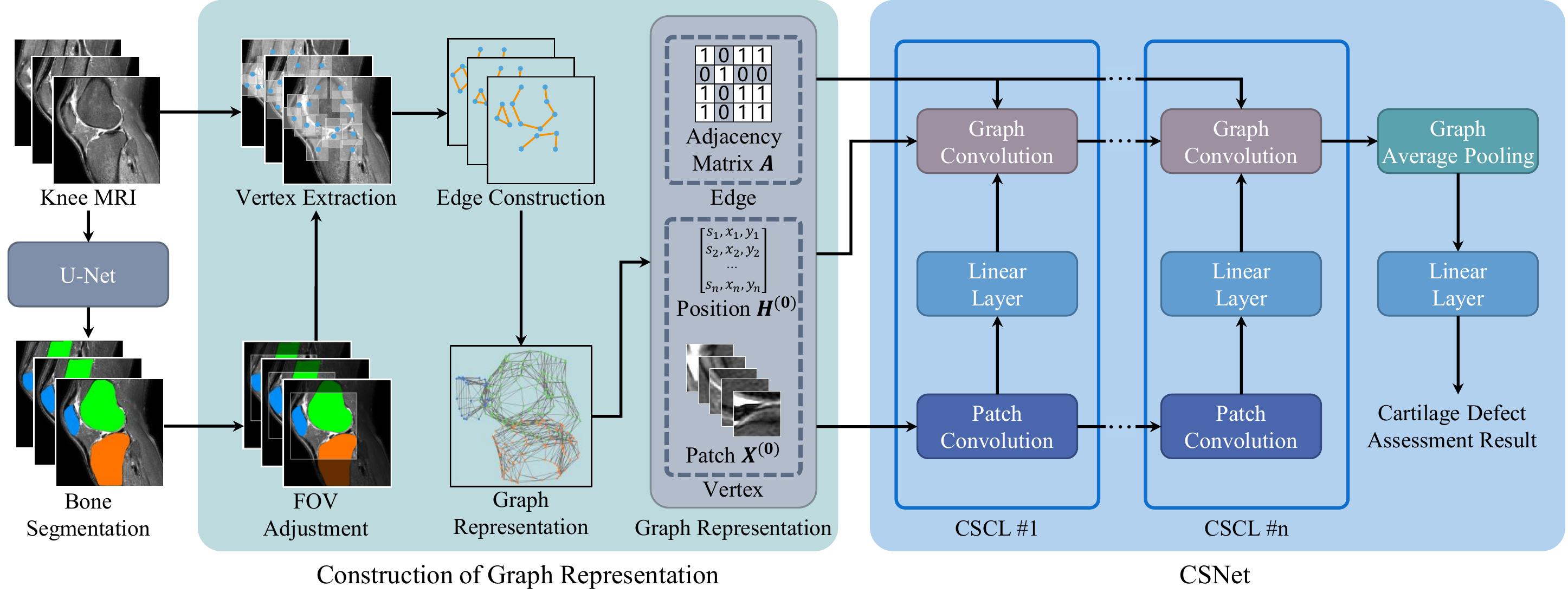}
\caption{
The framework of our proposed method consists of two major steps: (1) The graph construction, which converts heterogeneous knee MRI into graph representation; (2) The Cartilage Surface Network~(CSNet), which analyzes the graph representation and derives the defect assessment result.
}
\label{fig_overview}
\end{figure*}

Although these CNN-based researches are perceived as state-of-the-art tools for automatic knee cartilage defect assessment, several issues can hardly be ignored. 
\begin{enumerate}
    \item \textbf{Information Dilution}. For the subject-based and slice-based methods, they introduce large background into computation, which dilutes the portion of cartilage information. Knee cartilages are typically thin layers covering the heads of bones. As shown in Fig.~\ref{fig_intro}(a), only a very small portion of voxels in the form of curved shapes, other than the large background in knee MRI, should be focused in cartilage defect assessment. However, in conventional CNNs, the receptive field cannot be easily adjusted following the size and the shape of the target anatomy.
    \item \textbf{Structure Disidentification}. Patch-based methods reduce the irrelevant background by cropping patches along the cartilage. But the global shape information of knee cartilages is dropped out, too. In addition, the local appearance of individual patches at different positions of the cartilages is also highly diverse. We show patches with similar local appearance in Fig.~\ref{fig_intro}(b) and (c). Although (b) and (c) both have discontinuity of femoral cartilages, (b) indicates the femoral cartilage defect, and (c) is normal at the end zone of the femoral cartilage. Such confusion in local appearance can decrease the capability of patch-based methods in finding cartilage defects.
    \item \textbf{Heterogeneous Scanning Parameters}. It is common that the scanning parameters, including image size, in-slice spacing, inter-slice spacing, etc., are various for different subjects. A deep learning system, when applied to real clinical data of knee MRIs, is expected to handle such inhomogeneity automatically. However, the CNN-based methods typically require size-consistent inputs, which makes it difficult to generalize toward real clinical data.
    \item \textbf{Weak Interpretability}. Although existing CNN-based methods have the ability to provide an attention map to mark the possible areas of cartilage defects slice-by-slice, it is hard to interpret for 3D MRI. In terms of cartilage physiology, the defects should be detected and rendered in reference to the cartilage surface, which is mostly ignored in early methods.
\end{enumerate}

In this paper, we develop a novel method for knee cartilage defect assessment with two major parts to resolve the above issues: the cartilage graph representation to adapt to different MRI spacing and eliminate information dilution, and the cartilage surface network (CSNet) to enhance the cartilage structure identification and give strong interpretability. The pipeline of our framework is shown in Fig.~\ref{fig_overview}.
\begin{itemize}
    \item First, we propose a unified \textbf{graph representation} for all three knee cartilages (i.e., attached to femur, tibia, and patella) per subject. The graph auto-adaptively allocates vertices in adjacency to the cartilage surface, to account for knee MRIs with heterogeneous scanning parameters. Meanwhile, for individual vertices, the corresponding local patches, as well as their Cartesian coordinates, are extracted as their signatures to suppress information dilution. By connecting all vertices with the edges, the three cartilages are thus described by the unified graph representation, which captures not only global shapes but also the local appearance of the cartilages. 
    \item Second, we design the \textbf{CSNet} following a layer-by-layer design. In each layer, we adopt patch convolution to extract local appearance features from each vertex and then graph convolution to establish cross-talk among the vertices. Each layer is powered by the self-attention mechanism to boost the capability of modeling upon the cartilage graph. The final knee cartilage defect assessment is derived from pooling upon all vertices of the graph, while 3D visualization of the defect can also be conveniently attained to show the strong interpretability of the CSNet.
\end{itemize} 

The two major steps in our framework enhance the knee cartilage representation and improve defect assessment accordingly. In the experiments, we validate the efficiency of the cartilage representation and test the optimal setting of the CSNet, respectively. Then we show that the CSNet achieves superior performance in comparison to many state-of-the-art methods in the task of knee cartilage defect assessment. Moreover, we train and test the CSNet on the datasets with different inter-slice spacing, to demonstrate that our method is highly adaptive to the typical heterogeneity of clinical data. Finally, we show the attention map overlaid upon the mesh of the cartilage surface to verify the interpretability of our method.

\section{Related Work}
We first introduce the recent progress of deep learning for knee diseases in Section~\ref{subsec_dlinknee}. Then the non-Euclidean deep learning methods, related to the CSNet, are introduced in Section~\ref{subsec_nonculiedan}. Finally, the researches of network interpretability by visualizing the classification results are discussed in Section~\ref{sebsec_visualization}.

\subsection{Deep Learning in Image-Based Knee Disease Diagnosis}
\label{subsec_dlinknee}

Recently, deep learning methods have shown superior performance in many computer vision tasks~\cite{he2016deep, ouyang2020dual}.
This success has inspired many researchers and led them to focus on the computer-aided diagnosis~(CAD) of knee diseases to achieve promising results~\cite{bien2018deeplearningassisted, guida2021knee, pedoia20193d}. The major modalities of knee clinical image are radiograph and MRI.

Radiograph is a major target image modality in deep learning for CAD of knee disease. 
Knee X-rays are relatively cheap to obtain. The bone gap is clearly visible in 2D X-ray images, which provides evidence to conduct knee OA assessment. Thus many researchers use the Kellgren-Lawrence system as the criterion to grade knee OA based on the bone gap and achieve high classification accuracy~\cite{antony2016quantifying, tiulpin2018automatic, banymuhammad2019deep, swiecicki2021deep}. 

Compared to radiographs, MRIs have much better image quality for soft tissues and are superior for comprehensive observation of the knee joint, especially concerning cartilage defect, ACL lesion, and meniscus tear. However, the processing and understanding of knee MRIs are more challenging for deep learning: The curved cartilage structures, heterogeneous scanning parameters, and 3D data rather than 2D processing all pose difficulties for processing knee MRIs using conventional CNNs.

Researchers have used a variety of approaches to enable CNNs to operate on knee MRI and overcome these challenges. 
For example, to determine the feasibility of using deep learning to diagnose cartilage lesions, Liu et al.~\cite{liu2019fully} designed a CNN-based fully automated system. They particularly used the encoder of the segmentation CNN to guide the classification CNN and compared the automatic diagnosis performance of the system with musculoskeletal radiologists. 
Besides, Bien et al.~\cite{bien2018deeplearningassisted} made predictions from three views of knee MRI, and integrated all results with logistic regression for classification of knee abnormalities, anterior cruciate ligament~(ACL) tears, and meniscal tears. 
Merkely et al.~\cite{merkely2021improved} also trained three CNNs for individual views of knee MRIs separately, and derived image-specific saliency maps to visualize the decision-making progress of the deep networks. 

\subsection{Non-Euclidean Methods of Deep Learning}
\label{subsec_nonculiedan}
CNN can provide promising performance in deep learning tasks, as data in the Euclidean space and matrix form can be easily accelerated in GPU processing~\cite{zhang2012accelerated, hernandez2019using}. 
However, not all data is suitable for the representation in the Euclidean space ~\cite{bronsteinGeometricDeepLearning2017}, such as the knee cartilages focused on in this paper. There are two directions that are prevalent in tailing deep learning to compute in the non-Euclidean space: 
(1) Translating CNN architecture from the Euclidean space to the non-Euclidean space;
(2) Computing in the non-Euclidean space based on specially designed architectures such as graph convolutional network~(GCN) and Transformer. 

Many researchers adopt the well-established convolution-and-pooling scheme and work on the mesh-like data structure to address the non-Euclidean space problem~\cite{hanocka2019meshcnn, qi2017pointnet, zhao2021spherical}. Hanocka et al.~\cite{hanocka2019meshcnn} proposed MeshCNN to provide an operation similar to original CNN for 3D shapes in polygonal meshes. Motivated by PointNet~\cite{qi2017pointnet}, Hu et al.~\cite{hu2021subdivision} established SubdivNet for 3D triangular mesh classification and segmentation. Zhao et al.~\cite{zhao2021spherical} proposed SDU-Net for cortical surface by computing on the spherical mesh, and achieved the state-of-the-art performance in cortical thickness prediction.

Topology information of the data affects the network architectures significantly. Typical backbones include GCN\cite{chen2020simple}, hypergraph neural network~(HGNN)\cite{bai2021hypergraph} and Transformer~\cite{ying2021transformers}. Velickovic et al.~\cite{velickovic2018gat} presented graph attention network~(GAT), which introduced self-attention in the non-Euclidean space to the original GCN. Feng et al.~\cite{feng2019hypergraph} encoded the non-Euclidean data into a hypergraph structure for data representation and proved that HGNN is superior to GCN in the tasks with complex data relations. Yun et al.~\cite{yun2019graph} proposed the graph transformer network~(GTN), which could identify and reconstruct the edges between the nodes and achieved better accuracy in node classification than GCN.

\subsection{Interpretability of the Network}
\label{sebsec_visualization}
The interpretability of the network is related to the high-weighted features that influence the inference outputs. The weight spatially overlaid on the input can be seen as the network attention. Visualizing the attention of deep learning methods can help display the evidence for the networks to make a reasoning. 

Class activation mapping (CAM) is capable of visualizing the activated regions in CNN~\cite{zhou2016learning}. It uses a global average pooling layer and a fully connected (FC) layer to replace the multi-layer perceptron~(MLP), then uses the weights in the two layers to generate the attention map. While CAM can only show the attention of the last convolutional layer, Selvaraju et al.~\cite{selvaraju2017grad} extended GradCAM to probe the attention of more layers in the network. Further, Chattopadhay et al.~\cite{chattopadhay2018grad} proposed GradCAM++ accommodate the occurrence of the same yet multiple targets. 

These researches mentioned above focus on the attention maps generated in the Euclidean image space. To visualize the attention of data represented in the non-Euclidean space, Arslan et al.~\cite{arslanGraphSaliencyMaps2018} extended CAM to the GCN-based method for identifying the contributions of different brain regions in a gender classification task.

\section{Method}
\label{sec_method}
The framework of our proposed method is shown in Fig.~\ref{fig_overview}. Given sagittal T2 MRI that is popularly adopted to diagnose knee OA in clinical examination, our method assesses cartilage defects in two steps. First, we construct the graph representation, which captures the cartilages in both global shape and local appearance. Second, we design the CSNet, which jointly conducts patch convolution and graph convolution, to derive the final classification result. We further present the details to construct the graph representation of the cartilages in Section~\ref{subsec_graphrepresentation}, and to elaborate the CSNet in Section~\ref{subsec_CSNet}. We also provide more implementation details in Section~\ref{subsec_implementation_details}.

\subsection{Graph Representation of the Cartilages}
\label{subsec_graphrepresentation}
We define the graph $\bs{G}=\braces{\bs{V}, \bs{A}}$, which is composed of vertices $\bs{V}$ and edges $\bs{A}$, to represent the cartilages of each subject. The vertices cover the entire cartilage regions, while the edges connect individual vertices in the graph. 

The graph needs to be based on localizing the cartilages. To this end, we perform image segmentation in each MRI slice to label the bones of the femur, tibia, and patella by the U-Net~\cite{ronneberger2015u}. 
The vertices are extracted in reference to the bone labeling, instead of the cartilage segmentation. The reason is that cartilage segmentation can sometimes fail, concerning joint effusion and cruciate ligament, especially for OA patients. On the contrary, bone segmentation is much easier and more robust. As our goal here is to ensure that the vertices (and their patches) can cover the entire cartilage regions, bone labeling is a more suitable choice.

\subsubsection{Field-of-View Adjustment}
\label{subsubsec_FOVadj}
Based on the bone labeling, we adjust the field-of-view (FOV) for each MRI volume under consideration, before extracting the vertices and connecting them with edges. Particularly, we refer to the patella position and bone sizes in the knee joint to adjust the FOV \cite{suzuki2012vivo, ding2003sex}.
\begin{itemize} 
    \item The superior (upper) bound of the FOV is 9mm above the superior end of the patella, and the inferior bound is 100mm below the upper bound.
    \item The anterior bound is 3mm in front of the patella, and the posterior bound is 100mm behind the anterior bound. 
    \item The left bound is the first slice that includes any part of the bone label, and the right is the last slice with any bone. Note that we flip the left-right direction for all images of right knees. The subsequent computation thus focuses on the image space corresponding to the left knee.
\end{itemize}
The above procedure ensures all subjects have their cartilages included in the FOVs (c.f. Fig.~\ref{fig_overview}).

\subsubsection{Vertex Extraction}
\label{subsubsec_vertex}
We denote the set of voxels, which are on the surfaces of the labeled bones in the FOV, as $\bs{S}$. The vertex in $\bs{V}$ is drawn from $\bs{S}$, implying a much smaller size for $\bs{V}$ than $\bs{S}$. 

Initially, we represent the vertices as $\bs{V} = \{\bs{X}^{(0)}, \bs{H}^{(0)}\}$, where $\bs{X}^{(0)}$ is the collection for cartilage patches in knee MRI. In addition, we capture the Cartesian coordinates of the vertices as $\bs{H}^{(0)}$. For each vertex $\bs{v}_i$, it has the image patch $\bs{x}_i^{(0)} \in \mR^{p \times p}$ centered at its Cartesian coordinate $\bs{h}_i^{(0)} \in \mR^{3}$. Here, $p$ is the patch size. Since we consider clinical knee MRIs with high inter-slice spacing in this work, the patch $\bs{x}_i^{(0)}$ is in 2D rather than 3D. And we define $\bs{h}_i^{(0)} = (s, x, y)$, where $s$ is the index for the slice position, $x$ and $y$ are for the in-slice coordinates.
To ensure that all cartilages in each knee MRI are fully
covered by $\bs{V}$, we require two neighboring vertices in the same slice to have their patches partially overlapped, i.e., at a fixed level in terms of Intersection over Union~(IoU).
In this way, various numbers of vertices are automatically determined for each slice and each subject, to address high heterogeneity in imaging protocols and subjects.

\subsubsection{Edge Construction}
\label{subsubsec_edge}

\begin{figure}[!t]
\centering
\includegraphics[width=0.7\columnwidth]{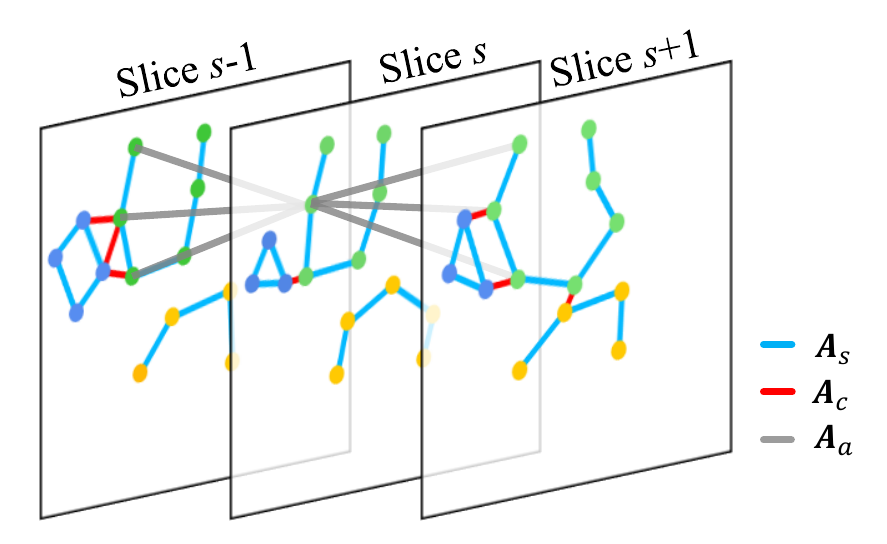}
\caption{
Edges connect vertices to ensure that the cartilage features can be exchanged on the cartilage graph in the CSNet. There are three types of edges: 
(1) $\bs{A}_s$, colored as blue, connects vertices along the bone surface in each knee MRI slice; 
(2) $\bs{A}_c$, colored as red, connects vertices in different cartilages of the same knee MRI slice; 
(3) $\bs{A}_a$, colored as grey, connects vertices in the same cartilage but different slices. 
\label{fig_edge}}
\end{figure}

We further establish edge connection for each knee MRI, which allows different vertices to exchange their features along the cartilage graph in the CSNet. The edge connection is stored by  the adjacency matrix $\bs{A}$
\begin{equation}
    \bs{A}[i,j]=1,~\text{if}~j \in \mathcal{N}_i, 
\end{equation}
where $\mathcal{N}_i$ is the set of connected vertices of $\bs{v}_i$. In this work, all the edges are undirected, or $\bs{A}[i,j]=\bs{A}[j,i]$. 

In clinical diagnosis, the continuity of cartilages is critical to identifying defects. Thus, we expect to examine the multi-scale local appearance of individual patches, and then exchange and ensemble them via the help from the edges of the graph. As shown in Fig.~\ref{fig_edge}, there are three types of edges: 

\begin{itemize}
    \item The continuity of the cartilage in a single slice is clearly visible to human experts. By connecting the relevant patches with the edges of $\bs{A}_s$, we can allow convolution among those patches that are adjacent with each other in the same slices. Such edges, which are blue colored in Fig.~\ref{fig_edge}, can be represented as:
    \begin{equation}
    \bs{A}_{s}[i,i+1]=1~,~\bs{v}_i, \bs{v}_{i \pm 1} \in \bs{V}_{l,s}.
    \end{equation}
    Here, $\bs{V}_{l ,s}$ indicates the vertices in the $l$-th cartilage ($l\in\{0,1,2\}$, corresponding to femur, tibia, and patella, respectively) and $s$-th slice. The vertices in $\bs{V}_{l ,s}$ are sorted along the $l$-th cartilage surface in the $s$-th knee MRI slice, such that $\bs{v}_i, \bs{v}_{i \pm 1} \in \bs{V}_{l ,s}$ are neighbors.
    
    \item Load bearing and friction between adjacent areas of different cartilages are highly related to cartilage defects in OA progression. To this end, we connect spatially close vertices in the same slice yet belonging to different cartilages (red-colored in Fig.~\ref{fig_edge}): 
    \begin{equation}
    \bs{A}_{c}[i,j]=1,~{\rm if}~{\left \|{\bs{v}_i}-{\bs{v}_j}\right \|} < D_c,\bs{v}_i\in \bs{V}_{l,s}, \bs{v}_j\in \bs{V}_{l \pm 1, s},
    \end{equation}
    where $\left \|{\bs{v}_i}-{\bs{v}_j}\right \|$ is the Euclidean distance between $\bs{v}_i$ and $\bs{v}_j$, $D_c$ is a threshold manually set to $p/2$.

    \item In most cases, cartilage defects can extend to more than one MRI slice. The continuity of the cartilage across multiple slices should also be considered. Thus, we define $\bs{A}_a$ as the edges passing features across adjacent slices:
    \begin{equation}
    \bs{A}_{a}[i,j]=1,~{\rm if}~{\left \|{\bs{v}_i}-{\bs{v}_j}\right \|} < D_a,\bs{v}_i\in \bs{V}_{l, s}, \bs{v}_j\in \bs{V}_{l, s \pm 1},
    \end{equation}
    where $D_a$ is a threshold. Considering the inter-slice spacing $t$ and the patch size $p$ (which controls the average distance of vertices in each slice), we set $D_a=0.8\sqrt{t^2+p^2}$. In this way, each $\bs{v}_i$ can be averagely connected to three vertices in each adjacent slice by $\bs{A}_a$, which is similar to the number of connected vertices in its own slice.

\end{itemize}
In the final, the edges can be represented by the overall adjacency matrix
\begin{equation}
\bs{A} = \bs{A}_s + \bs{A}_c + \bs{A}_a + \bs{I}.
\end{equation}
Particularly, each vertex is connected to itself, resulting in the identity matrix $\bs{I}$. Though $\bs{A}$ is a binary matrix initially, the edge connection strength $\bs{\hat{A}}$ will be learned in the CSNet.

\subsection{Cartilage Surface Network (CSNet)}
\label{subsec_CSNet}

\begin{figure}[!t]
\centering
\includegraphics[width=0.8\columnwidth]{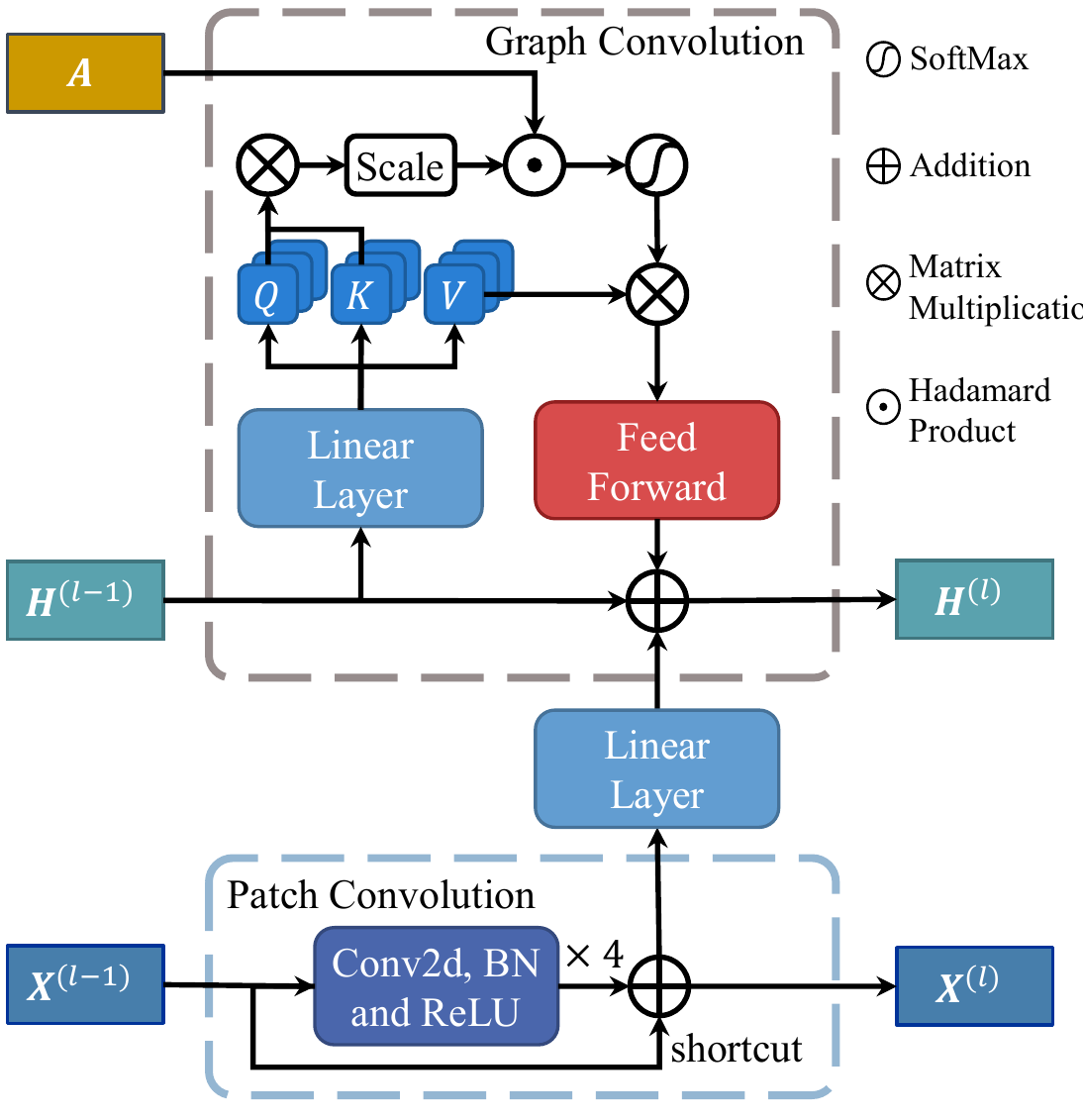}
\caption{
The CSCL consists of (1) patch convolution, including four stacks of 2D convolution, BN, and ReLU, to extract local appearance features of individual vertices, and (2) graph convolution, to exchange the features on the cartilage graph. 
}
\label{fig_layer}
\end{figure}

The above process transforms the cartilage regions in knee MRI from Euclidean space into the graph representation in the non-Euclidean space. To grade the knee cartilage defect, we develop the CSNet to extract local appearance features of individual vertices and exchange them on the cartilage graph by $L$ cartilage surface convolution layers (CSCLs).

The CSCL is illustrated in Fig.~\ref{fig_layer}. 
For the $l$-th CSCL, it receives $\bs{X}^{(l-1)}$ and $\bs{H}^{(l-1)}$ as the inputs.
Each CSCL consists of patch convolution and graph convolution. 
The patch convolution has stacked four blocks of 2D convolution, batch normalization (BN), and Rectified Linear Unit~(ReLU) with a shortcut for extracting the features $\bs{X}^{(l)}$. 
The graph convolution receives $\bs{H}^{(l-1)}$ from the previous layer and the linear projection of current $\bs{X}^{(l)}$. It then uses the self-attention mechanism to estimate the edge connection strength and updates the features of all vertices along with the graph. Note that the initial cartilage patches $\bs{X}^{(0)}$ and the positions $\bs{H}^{(0)}$ are generated following Section \ref{subsubsec_vertex}.
More details regarding patch convolution and graph convolution are described in Section~\ref{subsunsec_patchconv} and Section~\ref{subsunsec_graphconv}, respectively. 

After passing $L$ CSCLs, we acquire $\bs{H}^{(L)}$, upon which graph average pooling is applied. In this way, we derive $\bs{H}_g~\in~\mR^{C^{(L)}}$:
\begin{equation}
\bs{H}_g = \frac{\sum_{i}^{N} \bs{H}^{(L)}[i]}{N}, 
\end{equation}
where $N$ is the number of vertices in the cartilage graph.

Finally, the graph classification result $\bs{g}_{cls}$ can be calculated by a fully connected layer following
\begin{equation}
\bs{g}_{cls} = \bs{H}_{g} \bs{W}_{g},
\end{equation}
where $\bs{W}_{g}$ is for the weights of the FC layer. Finally, we can use the cross-entropy loss for the classification toward cartilage defect assessment.

\subsubsection{Patch Convolution} \label{subsunsec_patchconv}

Patch convolution plays the role of model local appearance features of individual patches in vertices. Concerning the small size of the patches in cartilage graph construction, patch convolution should be shallow. Particularly, we stack four blocks of 2D convolution, BN, and ReLU, plus a shortcut, in each CSCL layer.

Meanwhile, if cascading patch convolution of all CSCLs, we can add an extra pooling layer and an FC layer to reach a ResNet-like classifier. This classifier takes $\bs{X}^{(0)}$ as inputs and outputs the patch-level defect predictions. Based on this observation, we use labeled patches to pre-train this classifier and then separate individual blocks to the corresponding CSCLs.

\subsubsection{Graph Convolution} 
\label{subsunsec_graphconv}

In our method, the graph convolution allows the vertices to exchange their features $\bs{H}^{(l-1)}$ along the edges, and updates $\bs{H}^{(l)}$ with assistance of $\bs{X}^{(l)}$.

The classical GCN ~\cite{kipf2016semi} provides such a means. Specifically, GCN updates $\bs{H}^{(l)}$ in layer $l$ with the adjacency matrix $\bs{A}$ by 
\begin{equation}
\bs{H}^{(l)}=\sigma( \bs{D}^{-\frac{1}{2}} \bs{A} \bs{D}^{-\frac{1}{2}} \bs{H}^{(l-1)} \bs{W}^{(l)}).
\end{equation}
where 
$\bs{D}$ is the graph degree matrix ($\bs{D}[i,i]=\sum_j \bs{A}[i,j]$), 
$\bs{W}^{(l)}$ is the learnable weight of linear projection, 
and $\sigma$ is the feedforward function.

Considering the cartilage graph includes three different types of edges $\bs{A}_s$, $\bs{A}_c$ and $\bs{A}_a$, we further introduce the self-attention mechanism to graph convolution, to help distinguish the edge connection strength. First, a learnable weight matrix $\bs{W}_a$ is applied to $\bs{H}^{(l-1)}$ to compute the attention embedding of queries $\bs{Q}$, keys $\bs{K}$ and values $\bs{V}$ by
\begin{equation}
\bs{Q}, \bs{K}, \bs{V}=\bs{\hat{H}}^{(l-1)} \bs{W}_a,
\end{equation}
where $\bs{\hat{H}}^{(l-1)}$ is $\bs{H}^{(l-1)}$ after layer normalization. We adopt the multi-head strategy here, thus $\bs{Q}, \bs{K}, \bs{V} \in \mR^{N \times h \times d}$ with $h$ for the head number and $d$ for the length of embedding.

The self-attention mechanism applies the matrix product of the queries $\bs{Q}$ and the transposed keys $\bs{K}^T$ for the attention coefficient between any pair of vertices. Then we adjust the edge connection strength $\bs{\tilde{A}}$ by 

\begin{equation}
\bs{\tilde{A}} = \bs{A} \odot \frac{\bs{Q} {\bs{K}}^T} {\sqrt{d}}.
\end{equation}
Note that we only evaluate $\bs{Q}[i]$ with respect to $\bs{K}[j]$ for $j \in \mathcal{N}_i$ to avoid redundant computation. Further, we normalize $\bs{\tilde{A}}$ as
\begin{equation}
\bs{\hat{A}}[{i,j}] = \frac{\exp{(\bs{\tilde{A}}[{i,j}])}}{\sum_{k=0}^{N-1} \exp{(\bs{\tilde{A}}[{i,k}])}},
\end{equation}
where $\bs{\hat{A}}$ indicates the varying edge connection strength although the initial adjacency matrix $\bs{{A}}$ is fixed.

Finally, we derive $\bs{H}^{(l)}$ as the addition of the updated graph feature and the linear projection ($\bs{W}_p$) of $\bs{X}^{(l)}$ with a shortcut
\begin{equation}
\bs{H}^{(l)} = \sigma(\bs{\hat{A}} \bs{V}) + \bs{X}^{(l)} \bs{W}_p + \bs{H}^{(l-1)},
\end{equation}
with $\sigma$ representing the feedforward function. $\bs{H}^{(l)}$ will either be passed to next CSCL, or pooling toward the final assessment result.

\subsection{Implementation Details}
\label{subsec_implementation_details}
\subsubsection{Bone Segmentation}
\label{subsec_unet}
The cartilage graph is constructed in reference to the segmented femur, tibia, and patella bones. We use an in-house 2D U-Net~\cite{ronneberger2015u} to complete image segmentation here. We ask a radiologist to manually label 85 images as the training set for segmentation. The cross-entropy loss is adopted, with the Adam optimizer, to train the U-Net.

The segmentation of knee bones in the sagittal slices should ensure that the bone surfaces attached by cartilages are smooth and continuous. Thus we perform morphological operations to refine the output from the U-Net. Particularly, we use opening, region filling, and keeping maximal connected regions to ensure that each bone has a single piece of the label after segmentation.

\subsubsection{Batch Merging}
\label{subsec_graphbatch}

The heterogeneity of the knee MRIs brings in various graph sizes of different subjects. Thus, different subjects cannot be simply combined as a single batch directly. On the contrary, if each cartilage graph is treated as a batch, the training is time-consuming.

We adopt the deep graph library~\cite{wang2019dgl} to merge heterogeneous graphs in a batch. Specifically, for a certain batch of multiple subjects, we create an augmented adjacency matrix. For each subject, its graph and the corresponding adjacency matrix occupy a diagonal block in the adjacency matrix. As all subjects in the batch are non-overlapping in the augmented adjacency matrix, the graph of each subject can be perceived as a sub-graph behind the augmented matrix. Each sub-graph is disconnected from the others. In this way, all the graphs in the batch share the common (augmented) adjacency matrix, without cross-subject edges though. 

While this batch merging method does not add extra feature dimensions, each cartilage graph needs to be split from the batch before graph average pooling. The splitting operation is the inverse of the above batch merging.

\subsubsection{Multi-Level Supervision and Inference}
\label{subsec_trainingstragies}
Cartilage defect assessment can be done at the subject level, slice level, and patch level in the literature. In our method, we benefit from both subject-level and patch-level supervision. Specifically, we use patch-level supervision to pre-train a ResNet-like patch classifier, which consists of the patch convolution of all CSCLs, with an additional pooling layer and an FC layer. Then, we split them back to patch convolution of individual CSCLs. In this way, we can attain pre-training of the patch convolution for CSNet. After the pre-training is done, we only utilize subject-level labels to supervise the training of the CSNet.

Moreover, in inference, our method is unique to deliver the assessment results in all subject, slice, and patch levels. The subject-level output is naturally derived from the trained CSNet, whereas the slice-level result can be attained by treating the slice as a sub-graph and pooling upon all vertices of the slice only. The same FC layers are adopted in both subject-level and slice-level inference. The patch-level results can be acquired by adding an FC layer before the graph average pooling. Here the added FC layers are different from the pre-trained ResNet-like classifier, because the subject-level supervision after pre-training has altered the extracted patch features.

\section{Experiments}
In this section, we report the experimental results to validate the effectiveness of our proposed method on knee cartilage defect assessment.
First, we introduce the materials and experimental settings in Section~\ref{subsec_expsetting}.
Then, we test the effect of cartilage graph size on the CSNet computation, the contribution of the different types of edges to CSNet performance, the performance of the CSNet with different numbers of CSCLs as well as the performance of using alternative models for graph convolution, to justify the two major contributions of our method in Section~\ref{subsec_ablation} and find the optimal CSNet hyper-parameters.
After that, we compare the performance of our method with the state-of-the-art methods to prove our framework efficiency in Section~\ref{subsec_comparewithothers}. 
Moreover, in supplementary materials (due to page limit), we compare the performance of the CSNet with 3D CNN when training and testing on the datasets with different inter-slice spacing, to exhibit the superiority of our method in dealing with heterogeneous knee MRIs. We also visualize the attention of the CSNet to demonstrate its strong interpretability in localizing knee cartilage defects.

\subsection{Materials and Experimental Settings}
\label{subsec_expsetting}
The knee MRI dataset used in our experiments has been collected from Shanghai Jiao Tong University Affiliated Sixth People's Hospital. The subjects visited the hospital from 2011 to 2017, with ages ranging from 13 to 84 years old. There are in total 1205 images from a Philips Achieva 3.0T TX MRI scanner (Philips Healthcare, Best, Netherlands), using a T2-weighted sequence with fat suppression and fast spin echoes (repetition time: 2376ms; echo time: 62ms; slice thickness: 3mm; acquisition matrix: 292$\times$188; flip angle: 90°; bandwidth: 217.9kHz). Each subject contributes a single image, while the subjects with fractures, bone tumors, and atypical knee OA are excluded in recruiting.
\begin{table}[htbp]
\centering
\caption{Numbers of Subjects in Each Group of Our Clinical Knee MRI Dataset.}
\begin{tabular}{l|ll}
\hline\hline
 & Group & {\begin{tabular}[l]{@{}l@{}}\# Subjects\end{tabular}}
\\
\hline
\multirow{3}{*}{\begin{tabular}[l]{@{}l@{}}Grade of\\Cartilage Defect\end{tabular}} & Grade 0 & 518 \\ & Grade 1 & 370 \\ & Grade 2 & 317 \\
\hline
\multirow{3}{*}{\begin{tabular}[l]{@{}l@{}}Inter-Slice\\Spacing\end{tabular}} & = 3.3mm & 382 \\ & 3.6 -- 4.2mm & 43  \\ & = 4.5mm & 780 \\
\hline
\multirow{3}{*}{\begin{tabular}[l]{@{}l@{}}\# Sagittal Slices\\per Subject\end{tabular}} & = 18 Slices & 46 \\ & = 20 Slices & 770 \\ &  22 -- 29 Slices  & 389 \\
\hline
Total & &1205 \\
\hline\hline
\end{tabular}
\label{tab_data}
\end{table}

The original knee MRI parameters are shown in Table~\ref{tab_data}. And we follow Liu et al.~\cite{liu2018deep} and separate all images into three grades: Grade 0 (no defects, corresponding to the WORMS scores of 0 and 1), Grade 1 (mild defects, corresponding to the WORMS scores of 2, 3, and 4), and Grade 2 (severe defects, corresponding to the WORMS scores of 5 and 6). The assessment task is to determine the corresponding grade for each subject. The numbers of the subjects of Grade 0, Grade 1, and Grade 2 are 518, 370, and 317, respectively. 

The grading is labeled and reviewed by two radiologists. 
We collect labeling in the subject level, slice level, and patch level. 
In the subject level, the MRI scan comes with a grade as a whole. 
In the slice level, we ask the radiologists to score each sagittal MRI slice accordingly. 
And finally, in the patch level, all patches are cropped and radiologists grade all patches one by one. 
In summary, the subject label is the maximum of all slices per subject, and the slice label is the maximum of all patches with potential lesions in the slice.

All experiments are based on PyTorch1.8. The Adam optimizer is used and the weight decay is 1e-4. The learning rate is 3e-4 for our method, 3e-5 for Transformer and GAT, and 3e-4 for others. We have conducted five-fold cross-validation in each experiment and evaluated the three metrics: accuracy (ACC), recall (REC), and area under the curve (AUC).

\subsection{Contributions of Graph Representation and CSNet}
\label{subsec_ablation}
To justify the two major contributions of our method and find its optimal setting,
we conduct four studies to validate the graph representation (including the size of the graph and the types of edges) and the CSNet (including the number of CSCLs and the structure of graph convolution).

\subsubsection{Size of Graph}
\begin{table}[htbp]
\centering
\caption{Influence of Different Graph Sizes.}
\begin{tabular}{cc|c|ccc|c}
\hline\hline
\begin{tabular}[c]{@{}c@{}}Patch \\ Size\end{tabular} &\begin{tabular}[c]{@{}c@{}}In-Patch\\Spacing\end{tabular}  & 
$\overline{N_v}$ & ACC (\%) & REC (\%) & AUC (\%)  & FLOPs\\ \hline
32  & 0.303mm & 754 & 80.4$\pm$0.4&78.9$\pm$0.8&91.1$\pm$0.5&28.2G\\
64  & 0.303mm & 402 & \textbf{83.1$\pm$2.8}&\textbf{82.2$\pm$2.3}&\textbf{92.9$\pm$1.3}&35.5G\\
128 & 0.303mm & 214 & 79.7$\pm$2.1&78.9$\pm$3.0&92.4$\pm$1.3&52.2G\\
32  & 0.606mm & 492 & 78.6$\pm$1.4&77.7$\pm$0.9&90.1$\pm$0.5&19.3G\\
64  & 0.606mm & 263 & 79.4$\pm$0.9&77.8$\pm$1.4&90.2$\pm$1.6&24.5G\\
128 & 0.606mm & 138 & 68.9$\pm$2.5&66.7$\pm$1.6&83.4$\pm$1.3&35.5G\\
\hline\hline
\end{tabular}
\label{tab_patchsize}
\end{table}

To find the optimal cartilage graph representation, we examine how the graph size influences the 4-layer CSNet performance. 
As in Section~\ref{subsec_graphrepresentation}, one can adjust the number of vertices by controlling the initial patch size (e.g., a larger patch size reduces the number of vertices). In addition, considering the scanned MRIs, we set the recommended in-patch image spacing to be 0.303mm$\times$0.303mm. We can also increase the spacing (i.e., to 0.606mm$\times$0.606mm), such that the graph size can decrease. 

The experimental results show that both too large and too small graph sizes degrade the CSNet.
As shown in Row 2 of Table~\ref{tab_patchsize}, compared with Rows 1 and 3, the optimal graph size (average number of vertices $\overline{N_v}=402$) results in the best performance (ACC=83.1\%, REC=82.2\%, AUC=92.9\%). A too small patch may be limited in local awareness (c.f. the example in Fig.~\ref{fig_intro}(c)), while a too large patch may increase background appearance in the vertices of the graph and reduce the accuracy of the final assessment.

Meanwhile, given a fixed patch size, a coarser in-patch image spacing (e.g., from 0.303mm$\times$0.303mm in Row 2 to 0.606mm$\times$0.606mm in Row 5) can reduce the data granularity per vertex. Although the computation is saved in floating-point operations per second (FLOPs) as in Table~\ref{tab_patchsize}, the accuracy in assessing knee cartilage defects has been reduced significantly. As a summary, we adopt $p=64$ at 0.303mm$\times$0.303mm for our subsequent analysis.
\begin{table}[htbp]
\centering
\caption{Influence of Different Types of Edges in the Cartilage Graph.}
\begin{tabular}{ccc|ccc}
\hline \hline
\multicolumn{3}{c|}{Types of Edges} & \multicolumn{3}{c}{Subject-Level} \\ 
$\bs{A}_s$ & $\bs{A}_c$ & $\bs{A}_a$ & ACC (\%) & REC (\%) & AUC (\%) \\
\hline 
$\surd$  & $\surd$  & $\surd$  & \textbf{83.1$\pm$2.8}   & \textbf{82.2$\pm$2.3}    & \textbf{92.9$\pm$1.3} \\
$\surd$  & $\surd$  & $\times$ & 81.0$\pm$2.4    & 79.8$\pm$2.1    & 92.1$\pm$1.4 \\
$\surd$  & $\times$ & $\surd$  & 82.0$\pm$2.0    & 81.1$\pm$2.2    & 92.8$\pm$1.0 \\
$\times$ & $\surd$  & $\surd$  & 80.4$\pm$1.4    & 79.3$\pm$2.0    & 91.9$\pm$1.1 \\
$\surd$  & $\times$ & $\times$ & 80.7$\pm$2.5    & 79.0$\pm$2.9    & 92.1$\pm$1.5 \\
$\times$ & $\surd$  & $\times$ & 79.1$\pm$1.6    & 77.5$\pm$1.9    & 91.0$\pm$1.6 \\
$\times$ & $\times$ & $\surd$  & 78.2$\pm$1.9    & 77.6$\pm$1.3    & 90.0$\pm$0.8 \\
\hline \hline
\end{tabular}
\label{tab_edges} 
\end{table}
\begin{table}[htbp]
\centering
\caption{Influence of different numbers of CSCLs.}
\begin{tabular}{c|cccc}
\hline\hline
{\multirow{2}{*}{\# CSCLs}} & \multicolumn{3}{c}{Subject-Level}\\
  & ACC (\%)   & REC (\%)   & AUC (\%)   \\ 
  \hline
1 & 75.8$\pm$2.6 & 74.6$\pm$2.8 & 89.1$\pm$1.9 \\ 
2 & 78.5$\pm$2.8 & 77.5$\pm$3.3 & 90.8$\pm$1.6 \\ 
3 & 78.7$\pm$3.6 & 78.1$\pm$3.3 & 91.8$\pm$1.7 \\ 
4 & \textbf{83.1$\pm$2.8} & \textbf{82.2$\pm$2.3} & \textbf{92.9$\pm$1.3} \\ 
5 & 80.2$\pm$3.4 & 78.8$\pm$3.3 & 92.6$\pm$1.8 \\          
\hline\hline
\end{tabular}
\label{tab_csclsnumber}
\end{table}

\subsubsection{Types of Edges}
To prove and examine the contribution of the three types of edges connecting individual vertices in the graph ($\bs{A}_s$, $\bs{A}_c$ or $\bs{A}_a$ in Section~\ref{subsubsec_edge}), here we provide an ablation study by removing some of them to validate their necessity. 

As shown in Rows 1-4 of Table~\ref{tab_edges}, by removing $\bs{A}_s$ only, there is a bigger performance drop (e.g., ACC=80.4\% vs. 83.1\%) than removing any of the other two types of edges, indicating the most significant contribution from $\bs{A}_s$. Moreover, in Rows 5-7, when only one edge type is preserved, the $\bs{A}_s$-only case yields the best performance. The above finding is in line with our expectation that $\bs{A}_s$ represents the cartilage continuity in single sagittal knee MRI slices. Given the scanned clinical knee MRIs, it is the major source of evidence for radiologists to make a diagnosis.

Also, the absence of $\bs{A}_c$ and $\bs{A}_a$ decreases the performance of CSNet, e.g., by comparing Rows 1-3 in Table~\ref{tab_edges}. 
$\bs{A}_c$ and $\bs{A}_a$ are designed to make connections between local appearance where potential cartilage cross-talk happens, aiding CSNet to have long-range awareness and thus make a better inference. 
In summary, with all $\bs{A}_s$, $\bs{A}_c$ and $\bs{A}_a$, we can derive the best performance to assess knee cartilage defects. 

\subsubsection{Number of CSCLs}

The depth of the CSNet is controlled by the number of CSCLs, which should be considered for the balance between optimal performance and possible overfitting due to the graph complexity.
Thus we test different numbers of the stacked CSCLs. 
As shown in Table~\ref{tab_csclsnumber}, the best cartilage defect assessment result occurs when the number of CSCLs is 4. It can be seen that too few CSCLs may lack the ability to extract cartilage features effectively, while too many CSCLs may suffer from decreased performance as well. In all other experiments, we set the number of CSCLs to 4.

\subsubsection{Alternatives of Graph Convolution}
\begin{table}[htbp]
\centering
\caption{Comparisons of Graph Convolution of CSNet.}

\begin{tabular}{l|ccc}
\hline\hline

\multirow{2}{*}{} & \multicolumn{3}{c}{Subject-Level} \\
 & ACC(\%)           & REC(\%)          & AUC(\%)          \\ \hline
FC                              &75.2$\pm$2.4  &74.5$\pm$1.7  &91.1$\pm$1.9  \\ 
GCN~\cite{chen2020simple}       &80.9$\pm$2.0  &80.3$\pm$1.6  &92.8$\pm$0.7  \\ 
GAT~\cite{velickovic2018gat}    &81.7$\pm$2.6  &80.9$\pm$2.8  &92.4$\pm$1.2  \\ 
Transformer~\cite{vaswani2017attention}   &81.5$\pm$3.2  &80.0$\pm$3.9  &92.7$\pm$0.9  \\
CSNet      &\textbf{83.1$\pm$2.8} & \textbf{82.2$\pm$2.3}  & \textbf{92.9$\pm$1.3}  \\ 
\hline\hline
\end{tabular}
\label{tab_backbone}
\end{table}
We design the graph convolution with the self-attention mechanism, which computes edge connection strength and updates the cartilage features. To verify its effectiveness, we replace the graph convolution (from ``Linear Layer'' to ``Feed Forward'' in Fig.~\ref{fig_layer}) with alternative ways, yet keep the rest CSNet unchanged. The alternative implementations include the FC layer, original GCN layer~\cite{chen2020simple}, GAT layer~\cite{velickovic2018gat}, and Transformer layer~\cite{vaswani2017attention}. 

The comparisons are presented in Table~\ref{tab_backbone}. One can observe that our method achieves the best performance (ACC=83.1\%, REC=82.2\%, AUC=92.9\%). Moreover, FC has the worst classification performance due to its inability to exchange local appearance among the vertices. GAT performs better than GCN (ACC=81.7\% vs. 80.9\%), which is enhanced by the self-attention mechanism. Our method combines the self-attention mechanism with an individual linear layer and the graph structure to further improve the performance.

\subsection{Comparison with State-of-the-Art Methods}
\label{subsec_comparewithothers}

We compare our method with several literature reports to demonstrate that our proposed method can achieve state-of-the-art performance. 
Those compared methods include the works on knee MRIs for cartilage defect assessment: using the U-Net encoder to help patch-level classification~\cite{liu2018deep}, Dual-Consistency Mean-Teacher model (DC-MT) for slice-level classification~\cite{huo2020self}, and 3D-MedT~\cite{wang20213dmet} for 3D knee MRI classification at the subject level. 
Other methods that are not directly related to knee MRIs but have good performance in generalized visual applications are also compared, including Vision Transformer~(ViT)~\cite{dosovitskiy2020image}, Swin~Transformer~(SwinT)~\cite{liu2021swin}, ResNet~\cite{he2016deep} and 3D-ResNet~\cite{chen2019med3d}. 
While our method is capable of handling heterogeneous image data, the compared methods require standardized inputs in the Euclidean image space. Thus, we resample and pad all images to the size of 512$\times$512$\times$20 for them, with the in-slice spacing of 0.303mm$\times$0.303mm and inter-slice spacing unchanged.

\begin{table*}[thbp]
\centering
\caption{Comparing CSNet with Other Methods on Cartilage Defects Assessment.}
\begin{tabular}{l|ccc|ccc|ccc}
\hline\hline
\multirow{2}{*}{} &
  \multicolumn{3}{c|}{Subject-Level} &
  \multicolumn{3}{c|}{Slice-Level} & 
  \multicolumn{3}{c}{Patch-Level} \\ 
 & ACC(\%) &  REC(\%) & AUC(\%) & ACC(\%) & REC(\%) & AUC(\%) & ACC(\%) & REC(\%) & AUC(\%)
  \\ \hline
Liu et al.\cite{liu2019fully}      & -  & - & - & 87.2$\pm$1.0& 79.2$\pm$1.5& 94.5$\pm$0.5&86.0$\pm$1.5&76.9$\pm$1.3&94.3$\pm$0.5\\ 
DC-MT\cite{huo2020self}             & -  & - & - &84.6$\pm$1.2&76.6$\pm$1.8&93.3$\pm$0.6&84.5$\pm$0.6&72.9$\pm$1.6&91.8$\pm$0.7 \\ 
3D-MedT~\cite{wang20213dmet}                            &54.4$\pm$4.2&54.3$\pm$2.7&69.6$\pm$5.2& - & - & - & - & - & - \\ \hline
ViT\cite{dosovitskiy2020image}      & - & - & - &69.2$\pm$2.8&64.0$\pm$1.4&84.5$\pm$1.0&75.2$\pm$1.9&65.4$\pm$1.4&87.6$\pm$0.6\\ 
SwinT\cite{liu2021swin}             & - & - & - &71.6$\pm$2.2&64.7$\pm$0.5&84.9$\pm$0.5&80.5$\pm$1.8&71.6$\pm$1.5&90.9$\pm$0.7\\ 
3D-ResNet18~\cite{chen2019med3d}                         &67.7$\pm$4.5&67.0$\pm$3.7&81.6$\pm$3.4& - & - & - & - & - & - \\ 
3D-ResNet34~\cite{chen2019med3d}                       &66.5$\pm$3.2&65.5$\pm$3.7&81.0$\pm$3.4& - & - & - & - & - & - \\ 
3D-ResNet50~\cite{chen2019med3d}                         &63.4$\pm$2.5&62.8$\pm$1.7&78.3$\pm$1.5& - & - & - & - & - & - \\ \hline
ResNet18~\cite{he2016deep}                            & - & - & - &86.0$\pm$1.6&77.1$\pm$1.2&93.6$\pm$0.7&82.1$\pm$1.0&74.3$\pm$1.4&92.9$\pm$0.7\\ 
ResNet34~\cite{he2016deep}                            & - & - & - &85.4$\pm$1.4&76.4$\pm$2.2&93.5$\pm$0.4&84.0$\pm$1.7&74.0$\pm$1.4&92.4$\pm$0.6\\ 
ResNet101~\cite{he2016deep}                           & - & - & - &86.3$\pm$1.3&78.0$\pm$1.9&94.1$\pm$0.9&84.9$\pm$1.4&74.2$\pm$1.4&92.5$\pm$0.6\\ \hline
CSNet                               & \textbf{83.1$\pm$2.8}& \textbf{82.2$\pm$2.3}& \textbf{92.9$\pm$1.3}& \textbf{87.4$\pm$0.7} & \textbf{79.5$\pm$1.5} & \textbf{94.7$\pm$0.4} & \textbf{89.1$\pm$1.9}& \textbf{78.6$\pm$1.3}& \textbf{95.1$\pm$0.6}\\ \hline
\hline
\end{tabular}
\label{tab_compare_model}
\end{table*}

In Table~\ref{tab_compare_model}, our method achieves the best performance in diagnosing all three levels. 

In the subject level, 3D-ResNet ignores the curved knee cartilage shapes, thus suffering from cartilage information dilution and heterogeneous spacing. 3D-MedT is a variant of Transformer working on the 3D medical image directly, facing the same challenge. This is the reason for the low performance of the two methods. In contrast, our method has the best performance (ACC=83.1\%, REC=82.2\%, AUC=92.9\%). The graph representation divides the knee cartilage regions into local patches in vertices, allowing the CSNet to easily extract local appearance features and exchange them on the graph, and further avoids the information dilution and structure disidentification. 

The success of the CSNet at the slice level demonstrates the same fact - eliminating information dilution can bring in performance improvement for this task.
Additionally, the approach using the U-Net encoder~\cite{liu2019fully} demonstrates its strong performance (ACC=87.2\%), and ResNet~\cite{he2016deep} also gives good results (ACC=86.3\% for ResNet101). However, they both do not fully take into account the curved cartilage shape and the correlation of defects between the adjacent cartilages as the CSNet does. Thus their performance is lower than the CSNet (ACC=87.4\%, REC=79.5\%, AUC=94.7\%). Notably, the performance gap between other methods and the CSNet is reduced in the slice level, which we believe is due to the heterogeneity of inter-slice spacing being eliminated in this level.

At the patch level, the major negative impact is structure disidentification. Methods lacking priors on cartilage structures (e.g. ResNet~\cite{he2016deep}, DC-MT~\cite{huo2020self}, ViT~\cite{dosovitskiy2020image}, and SwinT~\cite{liu2021swin}) achieve worse results than Liu et al.~\cite{liu2019fully} that benefits from pre-trained bone segmentation. However, none of these methods allow local cartilage features to be exchanged along the cartilage surface and therefore lack the perception of the neighborhood. Our CSNet performs joint patch convolution and graph convolution, which achieves the best results.

Overall, our method outperforms the other methods at the subject, slice, and patch levels. This demonstrates the superiority of our framework for solving the challenges based on non-Euclidean cartilage graph representation and the CSNet.

\section{Discussion and Conclusion}

In this paper, we propose a framework to transform the knee cartilages into a unified graph representation and deal it with the CSNet. In this way, we effectively combine the cartilage shapes and appearance features for knee cartilage defect assessment at the subject, slice and patch levels. The experiments on the clinical knee MRIs demonstrate that our method not only achieves state-of-the-art performance but also shows its robustness in heterogeneous clinical knee MRIs.

Despite the above innovations and advantages, our proposed method still has some drawbacks that can be addressed in future research. First, we extract 2D patches along the bone surface as the initial local appearance signature due to the inter-slice spacing of the knee MRI. For other organs that are often scanned in 3D (e.g., brains or vessels), we can use 3D patches and eliminate the anisotropy of the graph representation. Second, although our method is robust to the heterogeneity of knee MRIs, we could not fully eliminate this effect. Better cartilage structure representation still needs to be explored. Further research directions include extending the graph representation and surface convolution to more organs and applications. We will also investigate the combination of the graph attention visualization for human-machine interaction in the intelligent diagnosis of medical images.
\bibliography{ref.bib}
\bibliographystyle{IEEEtran}
\end{document}